\documentclass[twocolumn,showpacs,preprintnumbers,amsmath,amssymb]{revtex4-1}

\usepackage{graphicx}
\usepackage{epstopdf}
\usepackage{float}
\usepackage{color}
\usepackage{amsmath}
\newcommand{\red}[1]{\protect\textcolor{red}{#1}}

\begin{document}

\title{First-order phase transition to a nonmagnetic ground state in nonsymmorphic NbCrP}

\author{Yoshiki Kuwata$^{1,2}$}
\author{Hisashi Kotegawa$^1$}
\author{Hideki Tou$^1$}
\author{Hisatomo Harima$^1$}
\author{Qing-Ping Ding$^2$}
\author{Keiki Takeda$^3$}
\author{Junichi Hayashi$^3$}
\author{Eiichi Matsuoka$^1$}
\author{Hitoshi Sugawara$^1$}
\author{Takahiro Sakurai$^4$}
\author{Hitoshi Ohta$^{1,5}$}
\author{Yuji Furukawa$^2$}
\affiliation{$^1$Department of Physics, Kobe University, Kobe 657-8501, Japan}
\affiliation{$^2$Ames laboratory, U.S. DOE and Department of Physics and Astronomy, Iowa State University, Ames, Iowa 50011, USA}
\affiliation{$^3$Muroran Institute of Technology, Muroran, Hokkaido 050-8585, Japan}
\affiliation{$^4$Research Facility Center for Science and Technology, Kobe University, Kobe, Hyogo 657-8501, Japan}
\affiliation{$^5$Molecular Photoscience Research Center, Kobe University, Kobe, Hyogo 657-8501, Japan}

\date{\today}

\begin{abstract}
We report the discovery of a first-order phase transition at around 125 K in NbCrP, 
which is the nonsymmorphic crystal with $Pnma$ space group. 
From the resistivity, magnetic susceptibility, and nuclear magnetic resonance measurements using crystals made by the Sn-flux method, the high-temperature (HT) phase is characterized to be metallic with a non-negligible magnetic anisotropy. 
The low-temperature (LT) phase is also found to be a nonmagnetic metallic state with a crystal of lower symmetry.
In the LT phase, the spin susceptibility is reduced by $\sim$30 \% from that in the HT phase, suggesting that the phase transition is triggered by the electronic instability. 
The possible origin of the phase transition in NbCrP is discussed based on the electronic structure by comparing it with those in other nonsymmorphic compounds RuP and RuAs.

\end{abstract}

\maketitle

\section{Introduction}
Phase transitions in metallic systems are generally driven by the cooperation of the Fermi surface property and electronic degrees of freedom such as spin, charge, and orbital.
Therefore, crystal symmetry can be a key ingredient for phase transition, particularly for a non-magnetic origin.
One of the examples is the charge density wave (CDW), which originates from the nesting property of the Fermi surface, and typically appears in low dimensional compounds.
Orbital ordering is another example where orbital degeneracies play an important role, associated with the lowering of crystal symmetry.
Recently the nonsymmorphicity of crystals has attracted much attention since it is also considered to affect phase transitions. 
In nonsymmorphic crystals, band degeneracies protected by glide or screw operations are realized along specific directions on the Brillouin Zone boundary \cite{burns}.
When the degenerate band is located near the Fermi level, the physical properties of materials will be strongly affected.
An example is the band Jahn-Teller effect which has been pointed out in RuP and RuAs as the origin of metal-insulator transitions at 270 and 250 K, respectively \cite{Hirai,Goto,Kotegawa}.
RuP and RuAs crystallize in three dimensional orthorhombic MnP-type structure with a space group of nonsymmorphic $Pnma$ which includes glide operations expressed by $n$ and $a$.
The transitions have been suggested to originate from the Fermi surface instability due to the degenerate flat bands \cite{Goto,Kotegawa}.
In the low-temperature nonmagnetic insulating phase, the superlattice formation was observed \cite{Hirai,Kotegawa}, although the Fermi surface nesting was not obvious due to the three dimensional nature of the materials \cite{Goto,Kotegawa}.
At present, although several experimental studies have uncovered the physical properties of RuP and RuAs \cite{Sato,Chen,Li,Nakajima,Ootsuki}, the origin of the nonmagnetic phase transitions, especially the reason for the superlattice formation, is still not well understood.
Nevertheless, nonsymmorphicity is thought to play an important role in producing the electronic instability responsible for these phase transitions.

\begin{figure}[tb]
\centering
\includegraphics[width=\linewidth]{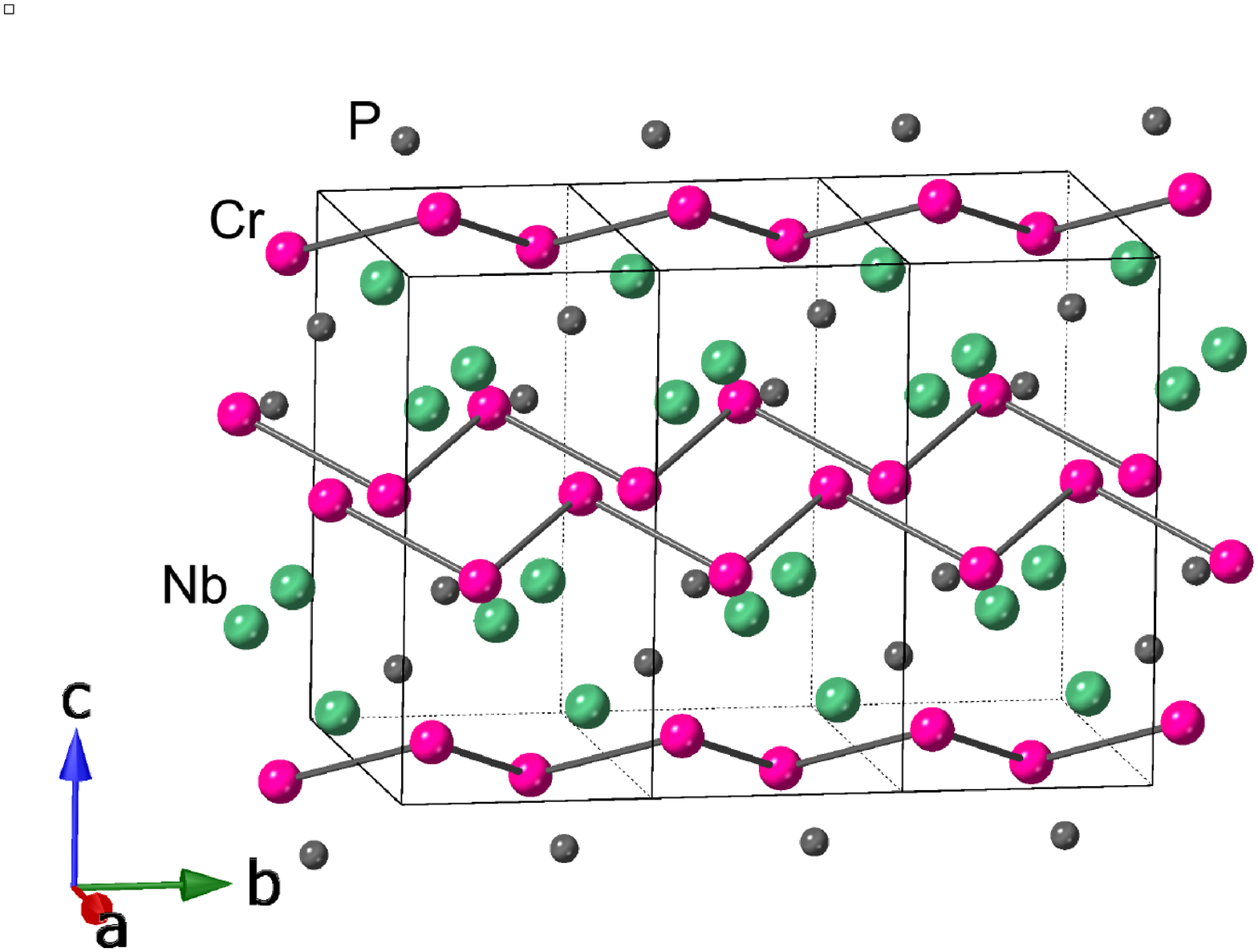}
\caption{\label{fig:CS}Crystal structure of NbCrP at room temperature (TiNiSi-type, space group $Pnma$).
The Cr ions form the zig-zag chains along the $b$ axis. All the ions occupy the equivalent sites in the $Pnma$ space group.
}
\end{figure}

Motivated by the nonmagnetic phase transitions observed in RuAs and RuP, we have been studying materials with MnP- and TiNiSi-type structures in order to investigate systematically the effects of nonsymmorphic symmetry on the physical properties of materials, especially focusing on phase transitions. 
The TiNiSi-type compound is a ternary system whose crystal symmetry is similar to that of MnP.
Both are in the $Pnma$ space group, and all the ions in the structures occupy the $4c$ sites which are locally non-centrosymmetric (here the local symmetry is $.m.$).
NbCrP is known to be a compound which crystallizes in the TiNiSi-type orthorhombic structure at room temperature \cite{Lomnytska}.
It is interesting to note that, similar to the case of RuP and RuAs where Ru ions form zigzag chains, the Cr atoms in NbCrP also form zigzag chains along the $b$ axis with a bond length of $\sim$2.62 \AA~as shown in Fig. \ref{fig:CS}. 
Since, to our knowledge, the physical properties of NbCrP have not been investigated so far, it is very interesting and important to investigate the electronic and magnetic properties of the compound.

In this paper, we carry out resistivity, magnetic susceptibility, and nuclear magnetic resonance (NMR) measurements of NbCrP.
From the experimental results, we found that a first-order phase transition to a nonmagnetic ground state occurs at $100-150$ K.
Clear suppressions of magnetic susceptibility and the nuclear spin-lattice relaxation rate in the low temperature phase suggest that the phase transition is accompanied by the reduction of the density of states at the Fermi level, $D(E_{\rm F})$, which originates from the instability of electronic states due to the characteristics of the crystal structure.
A comparison between NbCrP and RuAs is discussed based on calculated electronic band structures.

\section{Experimental Procedure}
Single crystals of NbCrP were attempted to grow with the Sn-flux method. Starting materials of Nb : Cr : P : Sn = $n$ : $n$ : 1 : $20n$ ($n=2-4$) were sealed in an evacuated quartz tube, which was heated up to $1050-1100$ $^\circ\mathrm{C}$, and then slowly cooled down to 600 $^\circ\mathrm{C}$ at a rate of $-5$ $^\circ\mathrm{C}$/h. After centrifugation, small needle-like crystals with a maximum length of about 0.5 mm were obtained. 
The obtained crystals are not single domain crystals but twinned crystals, as mentioned in the next section.
To crosscheck the physical properties of NbCrP, a polycrystalline sample was made by solid state reaction. Nb and Cr powder, and P flakes were sealed in an evacuated quartz tube with a stoichiometric ratio, and then heated 100 h at 1250$^\circ\mathrm{C}$, followed by 100 h at 1000 $^\circ\mathrm{C}$ after being ground. 
Single-crystal x-ray diffraction measurements were carried out by using a Rigaku Saturn724 diffractometer with a multi-layer mirror monochromated Mo-K$\alpha$ radiation. 
The data were collected by $\theta - 2\theta$ scans with a maximum $2\theta$ value of $\sim 62^{\circ}$. 
The program suite SHELX was used for the structure solution and least-squares refinement \cite{Sheldrick}.
PLATON was also used to check for missing symmetry elements in structures \cite{Spek}.
In addition to single crystal x-ray diffraction measurements which will be described in Sec. III, we also carried out x-ray diffraction measurements using polycrystalline samples.
The electrical resistivity of a crystal was measured using the four-probe method where electrical contacts of wires were made with the spot-welding method. 
Magnetic susceptibility $\chi$ was measured in a superconducting quantum interference device (SQUID) by utilizing a Magnetic Property Measurement System (MPMS : Quantum Design). 
NMR measurements were performed with the conventional pulsed spin-echo method using a phase coherent NMR spectrometer for the $^{31}$P (nuclear spin $I$ = 1/2 and 
gyromagnetic ratio $\gamma_{\rm N}$/2$\pi$ = 17.237 MHz/T) and $^{93}$Nb ($I = 9/2$, $\gamma_{\rm N}$/2$\pi$ = 10.405 MHz/T, and the quadrupole moment $Q$ = -0.22 barn) nuclei. 
The NMR spectra were obtained by recording the integrated spin-echo intensity by changing resonance frequency or magnetic field. 
$^{31}$P spin-lattice relaxation time $T_1$ was measured with the saturation recovery method. 
The observed recovery curves of the spin-echo intensity were well fit by a single exponential function as expected for a nuclear spin with $I$ = 1/2. 
For NMR measurements, we used two different types of samples. 
One was the many tiny crystals which are loosely packed into a sample case. 
The other sample was the oriented sample prepared by fixing the powdered crystals with stycast 1226 at room temperature under a magnetic field of $\sim$7.4 T, which allows the crystallites to be oriented along the applied magnetic field direction. 
From the NMR measurements, it turns out that, in the loosely packed sample, some of the crystals aligned along the applied magnetic field direction (that is, partially oriented sample).
Band structure calculations were performed with the full-potential linear augmented plane wave (FLAPW) method within the local density approximation (LDA).

\begin{figure}[tb]
\centering
\includegraphics[width=\linewidth]{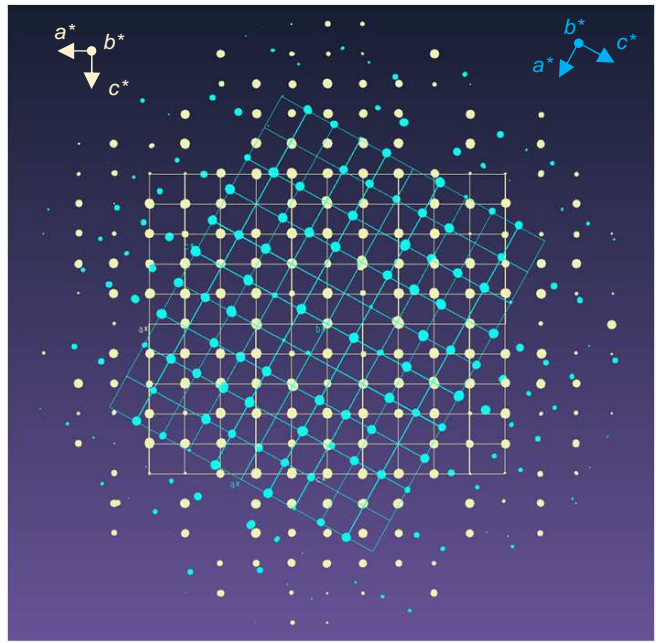}
\caption{\label{XRD_twin}X-ray diffraction pattern of a NbCrP twin crystal at room temperature (TiNiSi-type, Space group $Pnma$). 
The two sets of the reciprocal lattice patterns (blue and light yellow) are observed, showing that the crystal is twinned.
Corresponding reciprocal lattice vectors are shown by the arrows (the view is from the $b$-axis direction).
}
\end{figure}

\section{Crystal structure and characterization}

Crystallographic characterization of the crystals made using the Sn-flux method was performed by a single-crystal x-ray diffraction measurement. 
Figure \ref{XRD_twin} shows the x-ray diffraction pattern for a $0.05 \times 0.05 \times 0.04$ mm$^3$ crystal, where the superposition of two sets of reciprocal lattice appears. 
Each is consistent with the orthorhombic structure reported for the polycrystalline samples \cite{Lomnytska}.
This result indicates that the crystal is not single domain but twinned.
As shown in Fig.~2, we found an $\sim$60$^\circ$ difference in the $a$ (and $c$) axes between the two domains and the $b$ axes (corresponding to the needle direction of the crystal) are aligned in the two domains.
We checked several different crystals and always observed the twinned structure. 
The analysis shows that the deficiency at each site is estimated to be less than 1 \%, ensuring high quality of the crystal.
The structural parameters are summarized in Tables~I and II. 
The lattice parameters determined by the single crystal x-ray diffraction measurements are consistent with those for our polycrystalline sample within an error of $\pm$0.3\%, although they are slightly different from the previously reported values \cite{Lomnytska}.
We also confirmed there is no intrinsic difference in physical properties such as electrical resistivity and magnetic susceptibility between the twined and polycrystalline samples, although the data for the polycrystalline sample are not shown in this paper.
This suggests that the twinning of the crystal does not influence the electronic state of NbCrP.

\begin{table}[htb]
\centering
\caption{\label{table:structure}Crystallographic data at room temperature for NbCrP grown by the Sn-flux method.
}
\begin{tabular*}{\linewidth}{l@{\extracolsep{\fill}}cc}\hline \hline
Crystal systems & orthorhombic\\ \hline
Formula &NbCr$_{0.995}$P\\ \hline
Space Group & No. 62, $Pnma$, $D_{\rm 2h}^{\rm 16}$\\ \hline
$a$ (\AA)&6.2275(6)\\ \hline
$b$ (\AA)&3.5255(3)\\ \hline
$c$ (\AA)&7.3775(6)\\ \hline
Cell Volume (\AA$^3$) &161.973\\ \hline
$Z$ &4\\ \hline
\end{tabular*}
\end{table}

\begin{table}[htb]
\centering
\caption{\label{table:parameters}Structural parameters at room temperature of NbCrP grown by Sn-flux method.
}
\begin{tabular*}{\linewidth}{l@{\extracolsep{\fill}}ccccccc}\hline \hline
Site & Wyckoff& Site symmetry & $x$ & $y$ & $z$ & Occupancy \\ \hline
Nb & $4c$ & $.m.$ &0.5334&0.25&0.617 & 1 \\
Cr & $4c$ & $.m.$ &0.1400&0.25&0.4432 & 0.995 \\
P & $4c$ & $.m.$ &0.2650&0.25&0.1350 & 1 \\ \hline
\end{tabular*}
\end{table}

\section{Results and Discussion}
\subsection{Resistivity and magnetic susceptibility measurements}

\begin{figure} [b]
\centering
\includegraphics[width=\linewidth]{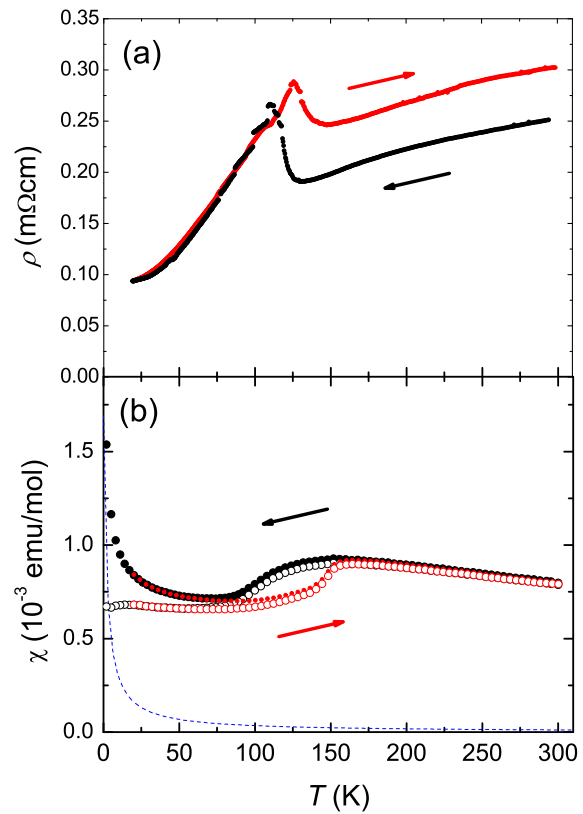}
\caption{\label{fig:bulk_property}Temperature dependences of (a) resistivity $\rho$ and (b) magnetic susceptibility $\chi$ of NbCrP. 
The $\chi$ was measured under 1 T for many small crystals (similar to powdered sample). Solid arrows indicate the direction of the thermal processes. 
Both $\rho$ and $\chi$ data clearly show a first-order phase transition at around $100-150$ K. 
The corrected $\chi$ by subtracting the Curie-Weiss contribution (blue dashed line) of paramagnetic impurity from observed $\chi$ is also shown by open symbols.}
\end{figure}

Figure \ref{fig:bulk_property}(a) shows the temperature dependence of $\rho$ below 300 K for NbCrP, where the current flows along the needle direction (the $b$ axis).
With decreasing temperature from room temperature, $\rho$ decreases and shows a clear increase at around 125 K due to a phase transition, and then decreases again below 100 K. 
These results clearly show that the low-temperature (LT) and high-temperature (HT) phases are metallic states. 
As shown in Fig. \ref{fig:bulk_property}(a), a clear hysteresis in $\rho$ is observed. 
Thus one can conclude that the phase transition is a first-order type. 
The different values of $\rho$ in the HT region between cooling and heating processes are most likely due to cracks in the crystal caused by the phase transition. 
In fact, we always observed cracks in the crystal after the thermal process, which changes the current path, producing the change in resistivity. 
Such behavior is often observed when a phase transition accompanies a large change in the volume of compounds \cite{Kotegawa_CrAs}.
Figure \ref{fig:bulk_property}(b) shows the temperature dependence of the magnetic susceptibility $\chi$ measured under a magnetic field of 1 T. 
In this measurement, many small crystals are tightly packed, so the crystals will be randomly oriented like powdered samples.
We were not able to measure $\chi$ using a piece of crystal because of the smallness of each crystal. 
$\chi$ increases gradually with decreasing temperature and starts to decrease around 125 K and then increases again at low temperatures. 
The upturns observed at low temperatures below $\sim$75 K are not intrinsic and are due to paramagnetic impurities as clearly evidenced by the nearly temperature independent behavior of the $^{31}$P NMR shift as will be described below. 
The open symbols in Fig. \ref{fig:bulk_property}(b) are the corrected $\chi$ by subtracting the Curie-Weiss contribution (shown by the blue dashed line) of the paramagnetic impurities from observed $\chi$.
The clear hysteresis is also observed in $\chi$, ensuring again a first-order phase transition. 
It is important to point out that the increase in $\rho$ and the suppression in $\chi$ just below the phase transition temperature indicates that the density of states at Fermi energy $D(E_{\rm F})$ is reduced in the LT phase from the HT phase.
As will be explained in the Sec. IV-E, our band-structure calculation gives a density of states at Fermi energy $D(E_{\rm F})$ of 193 states/Ry for the HT phase of NbCrP. This corresponds to a Pauli susceptibility of about $1.1\times 10^{-4}$ emu/mol. The observed susceptibility is about 7 times larger than this value, suggesting that it is somehow enhanced by the electron correlations.

\subsection{$^{31}$P-NMR measurements}

\begin{figure}[htb]
\centering
\includegraphics[width=\linewidth]{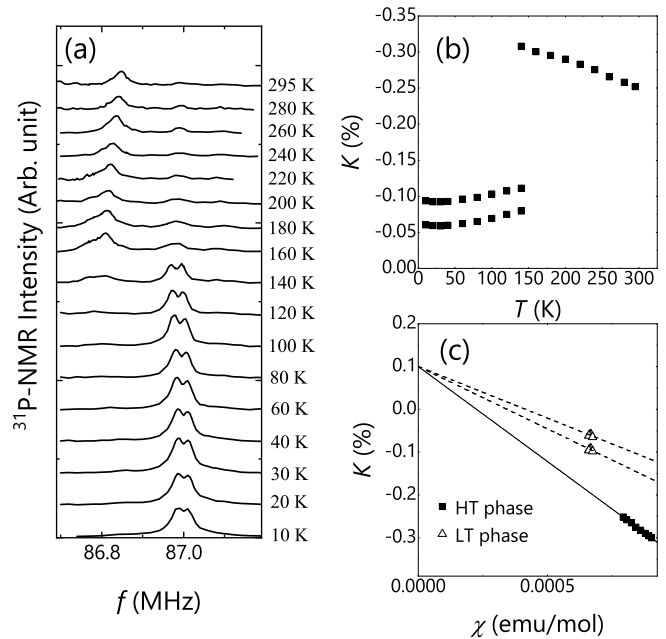}
\caption{\label{fig:P_loose}(a) Temperature dependence of $^{31}$P-NMR spectra for the loosely packed NbCrP sample, which is most likely partially oriented, measured at $H$ = 5.05 T. 
(b) Temperature dependence of the $^{31}$P-NMR Knight shift $K$ for the loosely packed powder sample. 
(c) $K$ vs. $\chi$ plot. The solid line is the fitting result for the HT phase. The dashed lines are drawn for the LT phase assuming that the $K$$_{\rm orb}$ is the same as that in the HT phase.} 
\end{figure}

Figure \ref{fig:P_loose}(a) shows the frequency-swept $^{31}$P-NMR spectra under $H$ = 5.05 T at various temperatures for the loosely packed single crystals of NbCrP.
In the LT phase, the double-peak line is observed at around 87 MHz. 
On the other hand, in the HT phase, a single peak with a slightly asymmetric shape is observed at a different position near 86.8 MHz. 
As will be shown below, it turns out that the crystals in the loosely packed sample are not totally random but partially oriented in the HT phase. 
Thus, the slightly asymmetric shape of the spectra observed in the HT phase is due to the anisotropic Knight shift and the partial orientation of the crystals.
We observe only the signal from the LT phase at low temperatures, giving clear evidence that the phase transition in NbCrP is bulk in nature. Since we measure the spectra from the lowest temperature with increasing temperature, we observe the signal from the LT phase up to 120 K.
Then, the NMR signals from the HT phase start to appear around 140 K where the coexistence of the two phases is clearly observed, consistent with the first-order phase transition.
With increasing temperature, the signal intensity of the NMR line from the HT phase increases, and that from the LT phase decreases, although the tiny signals from the remnant LT phase were observed even at room temperature.
The line width (full width at half maximum) in the LT phase is about $\Delta f \sim 80$ kHz ( $\sim 50$ Oe), which is comparable to that in the HT phase. 
In addition, we confirmed the spacing between the two peaks in the LT phase is proportional to the applied magnetic field, evidencing that the two peak structure originates from two inequivalent P sites, but not from static internal magnetic fields produced by magnetic ordering. 
Thus, we conclude the LT phase of NbCrP is nonmagnetic. 

\begin{figure}[htb]
\centering
\includegraphics[width=\linewidth]{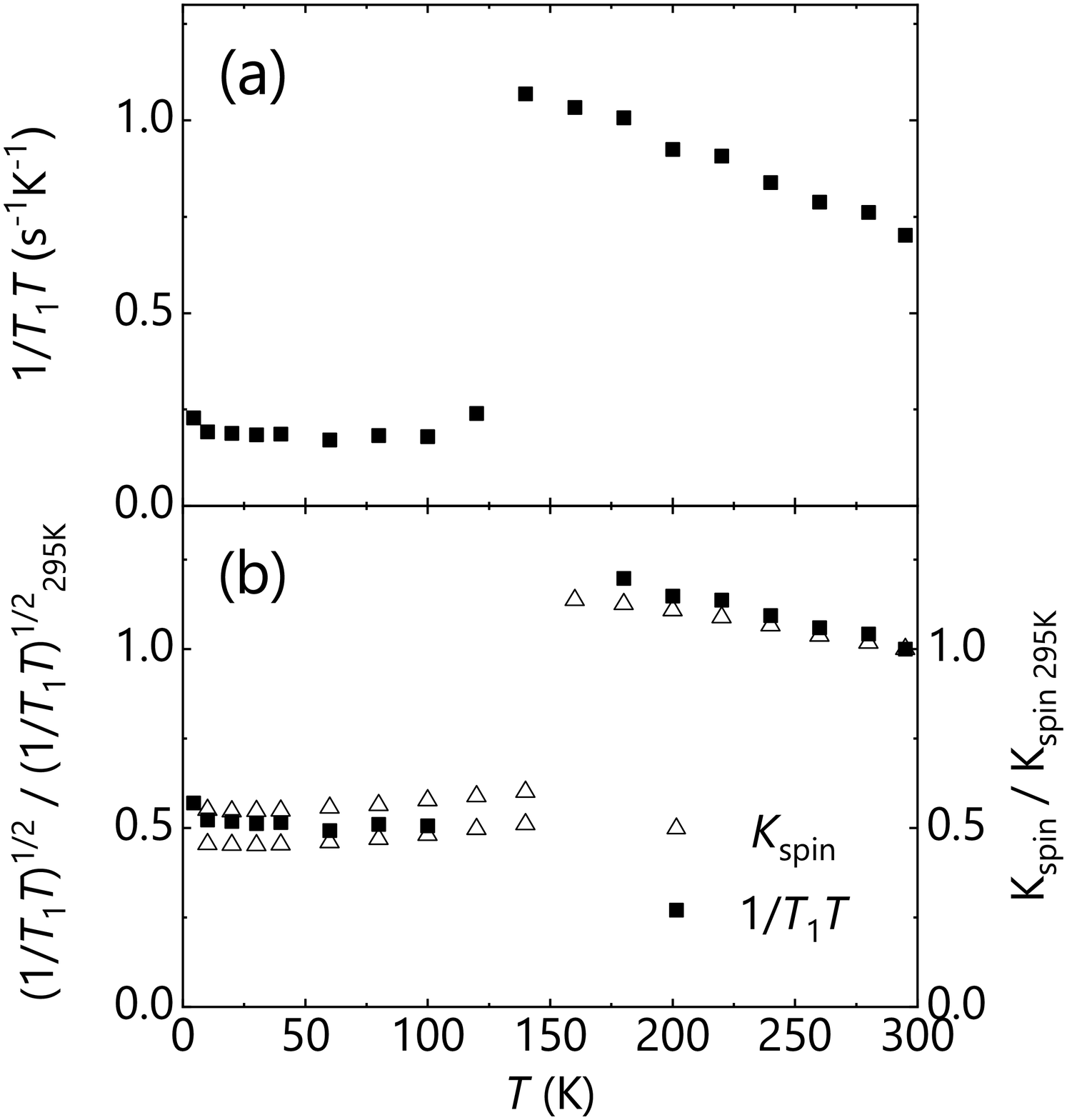}
\caption{\label{fig:P_T1}(a) Temperature dependence of the $^{31}$P $1/T_1T$. 
(b) Temperature dependence of $(1/T_1T)^{1/2}$ and $K_{\rm spin}$ normalized by the values at 295 K.
}
\end{figure}

Figure \ref{fig:P_loose} (b) shows the temperature dependence of the $^{31}$P-NMR Knight shift, $K$, which is determined by the peak positions of the spectra. 
The $K$ is negative in both LT and HT phases.
In the HT phase, $|K|$ increases with decreasing $T$.
On the other hand, $|K|$ is nearly independent of temperature with a smaller magnitude in the LT phase. 
This result indicates that the upturns in $\chi$ observed at low temperatures are not intrinsic and evidently arise from a small amount of paramagnetic impurities. 
The hyperfine coupling constant is estimated to be $A_{\rm HT}$ = $-2.5$ T/$\mu_{\rm B}$ for the HT phase from the slope of the $K-\chi$ plot [Fig.~4(c)] with the relation of $A_{\rm HT}$ = $\frac{N_{\rm A}\mu_{\rm B} K(T)}{\chi(T)}$, 
where ${N_{\rm A}}$ is Avogadro's number. Here we used the corrected $\chi$. 
From the vertical-axis intercept of the $K-\chi$ plot, the temperature independent part of $K$ is estimated to be $K_{\rm orb}=0.100 \pm 0.007$ \%.
We also plot the data for the LT phase in Fig.~4(c), where only the data below 80 K are used. 
As shown, the data for the LT phase are not on the line for the HT phase data. 
This indicates that the hyperfine coupling in the LT phase is different and is changed by the phase transition. 
It is also possible to have a change in the orbital part of the Knight shift, however, as we will see below, $K_{\rm orb}$ does not seem to be much changed in the LT phase. 
Since the spin part in $|K|$ is proportional to the density of states at the Fermi energy $D(E_F)$, the abrupt change in $K$ at around 150 K indicates that $D(E_{\rm F})$ decreases in the LT phase. 

The reduction in $D(E_{\rm F}$) in the LT phase is also observed in the temperature dependence of $1/T_1T$ shown in Fig. \ref{fig:P_T1}(a), where $1/T_1T$ is nearly independent of $T$ in the LT phase, while it depends on $T$ in the HT phase.
As the $1/T_1T$ is proportional to the square of $D(E_{\rm F})$, 
the large suppression of 1/$T_1T$ in the LT phase compared with that in the HT phase indicates the reduction of $D(E_{\rm F}$) in the LT phase, consistent with the $K$ data.
Figure \ref{fig:P_T1}(b) show the temperature dependence of $(1/T_1T)^{1/2}$, together with that of the spin part of the Knight shift $K_{\rm spin} (= K - K_{\rm orb}$) which is also proportional to $D(E_{\rm F})$.
Here the values of $(1/T_1T)^{1/2}$ and $K_{\rm spin}$ are normalized by the corresponding values at 295 K for each. 
By taking $K_{\rm orb}$ = 0.1 \% estimated from the $K-\chi$ plot for the HT phase, we found that the temperature dependence of $(1/T_1T)^{1/2}$ and $K_{\rm spin}$ are well scaled. 
This indicates that the $K_{\rm orb}$ does not change much in the LT phase.
Assuming $K_{\rm orb} = 0.1$ \%, we estimate the hyperfine coupling constants for the two P sites in the LT phase to be $A_{\rm LT} = -1.6$ and $-1.4$ T/$\mu_{\rm B}$ from the slope of the dashed line shown in Figure \ref{fig:P_loose}(c).
Utilizing the values of $A_{\rm HT}$ and the averaged $A_{\rm LT}$ deduced from the $K-\chi$ plot, we can estimate an $\sim$30 \% reduction of the spin susceptibility for the LT phase from the HT phase by the phase transition.

\subsection{$^{31}$P-NMR measurements for the oriented sample}

\begin{figure}[b]
\centering
\includegraphics[width=\linewidth]{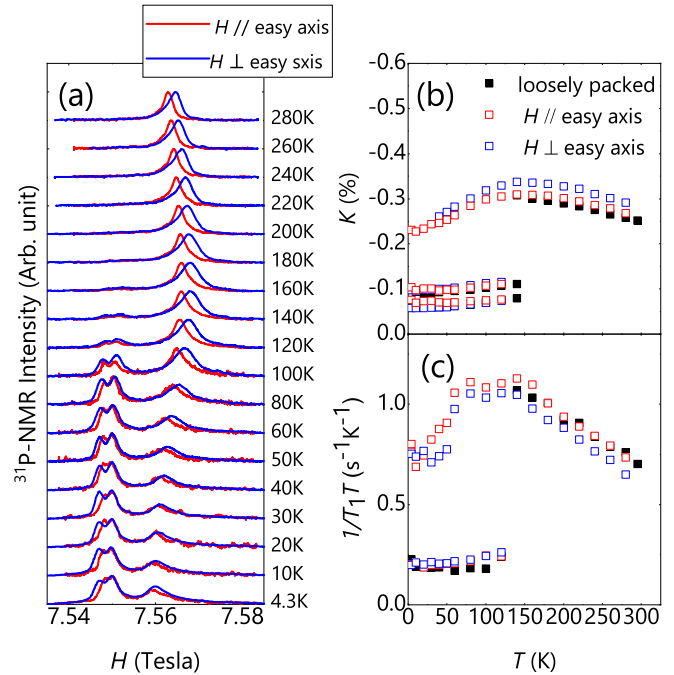}
\caption{\label{fig:P_orient}(a) Temperature dependence of field-swept $^{31}$P-NMR spectra ($f$ = 130 MHz) for oriented samples measured under magnetic fields parallel to the easy axis (red curves) and perpendicular to the easy axis (blue curves).
(b) Temperature dependence of the $^{31}$P-NMR Knight shifts for the oriented samples for the two magnetic field directions. 
(c) Temperature dependence of the $^{31}$P $1/T_1T$. }
\end{figure}

To investigate further details of magnetic correlations in NbCrP, we performed NMR measurements for the oriented sample.
Here, the powdered crystals were oriented under a magnetic field and fixed with Stycast 1266 at room temperature.
Figure \ref{fig:P_orient}(a) shows the temperature dependence of field-swept $^{31}$P-NMR spectra measured for the oriented sample under magnetic fields parallel $H_{||}$ (red) and perpendicular $H_{\perp}$ (blue) to the oriented direction.
The observation of the different spectra for the two magnetic field directions clearly indicates that most crystals in the oriented sample are aligned along the magnetic field direction corresponding to the magnetic easy axis of NbCrP, revealing a non-negligible magnetic anisotropy in the HT phase.
The double-peak structure of the spectra in the LT phase is similar to the case of the loosely packed sample, but it is interesting to point out that we observed the signals from the HT phase down to 4.3 K.
Since we do not observe such signals in the loosely packed sample, we consider that it is most likely due to inhomogeneous local pressure produced by the different thermal expansion properties between NbCrP and epoxy (Stycast 1266).
This indicates that the phase transition in NbCrP is quite sensitive to pressure and low pressure will suppress the phase transition.
Owing to this, we are able to track the temperature dependence of the Knight shift and 1/$T_1T$ for the HT phase in a wide temperature range down to 4.3 K.
Figure \ref{fig:P_orient} (b) shows the temperature dependence of the Knight shift parallel to the easy axis ($K_{||}$) and perpendicular to the easy axis ($K_{\perp}$).
It is noted that the data for the loosely packed sample are almost overlapped with $K_{||}$, implying that it is partially oriented in the HT phase.
$K_{\perp}$ and $K_{||}$ for the HT phase exhibit broad maxima around 125 K.
This indicates that the whole temperature dependence of $\chi$ for the HT phase exhibits a broad maximum around 125 K, which can be obtained only with the oriented sample fixed by the epoxy.
Similar broad maxima are also observed in the temperature dependence of 1/$T_1T$ as shown in Fig. \ref{fig:P_orient}(c). 
It is interesting to point out that the $K$ and 1/$T_1T$ are anisotropic in the HT phase, while no anisotropy is observed in the LT phase. 
This means that the bands inducing magnetic anisotropy, which are most likely Cr-$3d$ bands, are responsible for the phase transition to the nonmagnetic ground state.

The characteristic temperature dependence of $\chi$ and 1/$T_1T$ is different from that of simple metals.
The enhancement of $\chi$ from the estimation based on the band  structure calculation suggests the presence of moderate magnetic correlations.
However, the good scaling between $(1/T_1T)^{1/2}$ and $K_{\rm spin}$, shown in Fig.~5(b), contradicts either ferromagnetic (FM) or antiferromagnetic (AFM) correlations developing in NbCrP.
Therefore the temperature dependence of $\chi$ and 1/$T_1T$ can be interpreted in two ways.
One is the coexistence of AFM and FM correlations, which was reported in SrCo$_2$As$_2$ \cite{Pandey2013,Wiecki2015,Li2019}. 
Another is a contribution from a sharp peak in the density of states near the Fermi energy, which has been discussed in the skutterudite compounds CePt$_4$Ge$_{12}$ \cite{Toda2008} and SrFe$_4$As$_{12}$ \cite{Ding2018}.

As is known, it is useful to estimate the quantity $T_1TK_{\rm spin}^2$ to discuss the magnetic correlations \cite{Moriya1963,Narath1968}.
The so-called Korringa ratio 
${\cal K}$$(\alpha)\equiv$ $\frac{ z\cal S}{T_1TK_{\rm spin}^2}$
is unity for uncorrelated metals and greater (smaller) than unity for AFM (FM) correlated metals.
Here ${\cal S}$ = $\frac{\hbar}{4\pi k_{\rm B}} \left(\frac{\gamma_{\rm e}}{\gamma_{\rm N}}\right)^2$ where $\gamma_{\rm e}$ and $\gamma_{\rm N}$ are the electron and nuclear gyromagnetic ratios, respectively, and $z = 4$ is the number of nearest neighbor Cr irons with respect to a P atom.
Using the values of 1/$T_1T$ and $K_{\rm spin}$ in the HT phase for the loosely packed sample, we found that ${\cal K}$$(\alpha)$ is $\sim$ 0.8 close to unity.
This suggests that  either the magnetic correlations are weak in NbCrP or both AFM and FM correlations coexist, consistent with the above interpretation of the temperature dependences of  $\chi$ and $1/T_1T$.
To elucidate the magnetic correlations in the HT phase as well as the LT phase of NbCrP, further experiments, especially neutron diffraction measurements, are necessary.
In any case, the NMR results clearly suggest that the magnetic correlations in NbCrP are not dominated by the AFM correlations.
One of the possibilities for the phase transition to the nonmagnetic ground state in NbCrP is a dimerization of the Cr spin on the Cr zigzag chain through the AFM interaction, as discussed for RuP \cite{Li}, although a full-gap is not opened in NbCrP.
The present NMR results seem to contradict this scenario.

\subsection{$^{93}$Nb-NMR measurements for the oriented sample}

\begin{figure}[htb]
\centering
\includegraphics[width=\linewidth]{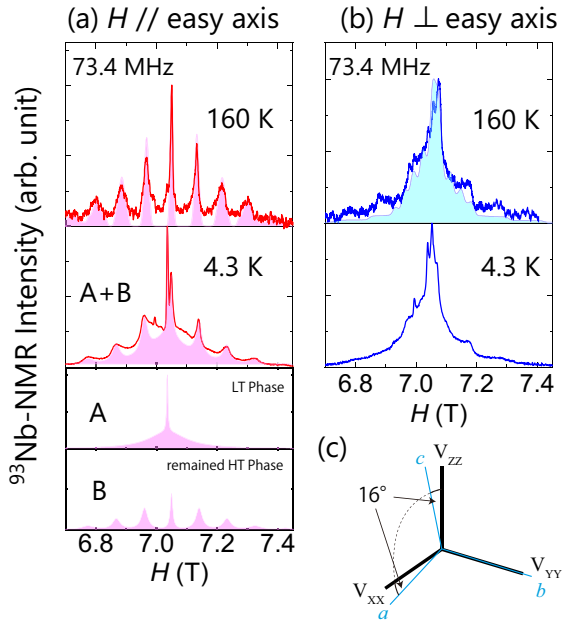}
\caption{ Typical field-swept $^{93}$Nb-NMR spectra measured using the oriented sample in the LT phase ($T$ = 4.3 K) and the HT phase ($T$ = 160 K) for (a) $H_{\rm ||}$ and (b) $H$$_{\rm \perp}$.
The colored areas in the figures show the simulated results.
(c) The schematic view shows the directions for principal axes of the EFG with respect to the crystalline axes obtained from the electronic band structure calculation. }
\label{fig:Nb-NMR}
\end{figure}

In order to determine the magnetic easy axis and also to gain more insights into the magnetic and electronic properties of NbCrP, we have carried out $^{93}$Nb-NMR spectrum measurements.
Figures \ \ref{fig:Nb-NMR}(a) and \ref{fig:Nb-NMR}(b) show the typical field-swept $^{93}$Nb-NMR spectra observed using the oriented sample in the LT ($T$ = 4.3 K) and the HT ($T$ = 160 K) phases for $H_{||}$ and $H_{\perp}$, respectively. 
The well separated lines observed in the HT phase for $H_{||}$ [Fig. \ref{fig:Nb-NMR}(a)] indicate the degree of the orientation of the oriented sample is relatively high. 
The spectrum is well explained by the following nuclear spin Hamiltonian which produces a spectrum with a central transition line flanked by four satellite peaks on both sides for $I$ = 9/2, 
\begin{eqnarray*} 
{\cal H} = {\cal -}\gamma_{\rm N}\hbar \mbox{\boldmath $I$}\cdot \mbox{\boldmath $H$}+ \frac{h \nu_{\rm Q}}{6} [3I_{Z}^{2}-I^2+ \frac{1}{2}\eta(I_+^2 +I_-^2)],
\end{eqnarray*}
where $h$ is Planck's constant, and $\nu_{\rm Q}$ is the nuclear quadrupole frequency defined by 
\begin{eqnarray*} 
\nu_{\rm Q} = \frac{3e^2QV_{ZZ}}{2I(2I-1)h} = \frac{e^2QV_{ZZ}}{24h} 
\end{eqnarray*}
for $I$ = 9/2, where $Q$ is the electric quadrupole moment of the Nb nucleus, $V_{ZZ}$ is the maximum electric field gradient (EFG) at the Nb site, and $\eta$, defined by $(V_{XX} -V_{YY}) /V_{ZZ}$ (here $|V_{ZZ}| \ge |V_{YY}| \ge |V_{XX}|$) is the asymmetry parameter of the EFG \cite{Slichterbook}.
In order to determine the principal axes of the EFG at the Nb site for the HT phase, we have carried out the electronic structure band calculation using the lattice parameters shown in Tables I and II.
As shown in Fig. \ref{fig:Nb-NMR}(c), the maximum principal axis ($V_{\rm ZZ}$) is found to be tilted from the $c$ axis to the $a$ axis by 16$^{\circ}$, and $V_{YY}$ is directed along the $b$ axis.

The light pink areas shown at the top panel of Fig. \ref{fig:Nb-NMR}(a) are the calculated $^{93}$Nb NMR spectrum using the parameters $\nu_Q=1.2$ MHz, $\eta=0.44$, $\theta = 90^\circ$ and $\phi = 0^\circ$.
Here $\theta$ and $\phi$ are the polar and azimuthal angles between the $V_{ \rm ZZ}$ and the direction of $H$, respectively.
As shown by the red curves, the observed $^{93}$Nb spectrum was well reproduced by the calculated one.
From the analysis of the spectrum, it turns out that the magnetic easy axis for the HT phase is along the $b$ axis corresponding to the direction of the zigzag chains formed by the Cr ions (see, Fig. 1). 
On the other hand, one expects a two dimensional powder pattern for $H_{\perp}$ since the crystallites are aligned only along the $b$ axis and no order is expected in the $ac$ plane in the oriented sample.
In fact, we observed the powder pattern $^{93}$Nb-NMR spectrum as shown in Fig. \ref{fig:Nb-NMR}(b), which is also reasonably reproduced by the calculated spectrum (shown in light blue) using the same set of parameters with appropriate anisotropy in $K$.
Thus one can conclude that the magnetic easy axis points along the $b$ axis.

At low temperatures, the spectrum for $H_{||}$ shows a drastic change although such an obvious change in the spectrum is not observed for $H_{\perp}$, as shown in Figs. \ \ref{fig:Nb-NMR}(a) and \ \ref{fig:Nb-NMR}(b).
Here one needs to be careful in interpreting the spectrum. 
As described in Sec. IV-C, we found that the HT phase remains even at low temperatures due to the effects of the epoxy.
From the detailed measurements of the temperature dependence of the spectrum, we figured out that the spectrum is well explained by a superposition of two spectra from the HT and LT phases, as shown in the case of $H_{||}$.
The remaining spectrum from the HT phase produces clear quadrupolar split lines depicted as B in the bottom panel of Fig. \ \ref{fig:Nb-NMR}(a). 
On the other hand, the spectrum from the LT phase shows no clear splitting of the satellite lines whereas the central transition line is relatively sharp as shown by A in the third panel of Fig. \ref{fig:Nb-NMR}.
As shown by the light pink area in the second panel of Fig. \ref{fig:Nb-NMR}, the superposition of the calculated spectra A and B reproduces reasonably the observed spectrum. 
The sharp central line for A suggests no significant internal field at the Nb site, consistent with the nonmagnetic state of the LT phase.
The broadened satellite lines indicate the large distributions in $\nu_{\rm Q}$ and/or in the directions of the principal axes of the EFG, suggesting the appearance of multiple Nb sites in the LT phase.
Together with the result for the $^{31}$P NMR where the single P site in the HT phase splits into two P sites in the LT phase, these results suggest that the lowering of the crystal symmetry occurs in the LT phase. 


\subsection{Electronic band structure calculation}

\begin{figure}[htb]
\centering
\includegraphics[width=1.0\linewidth]{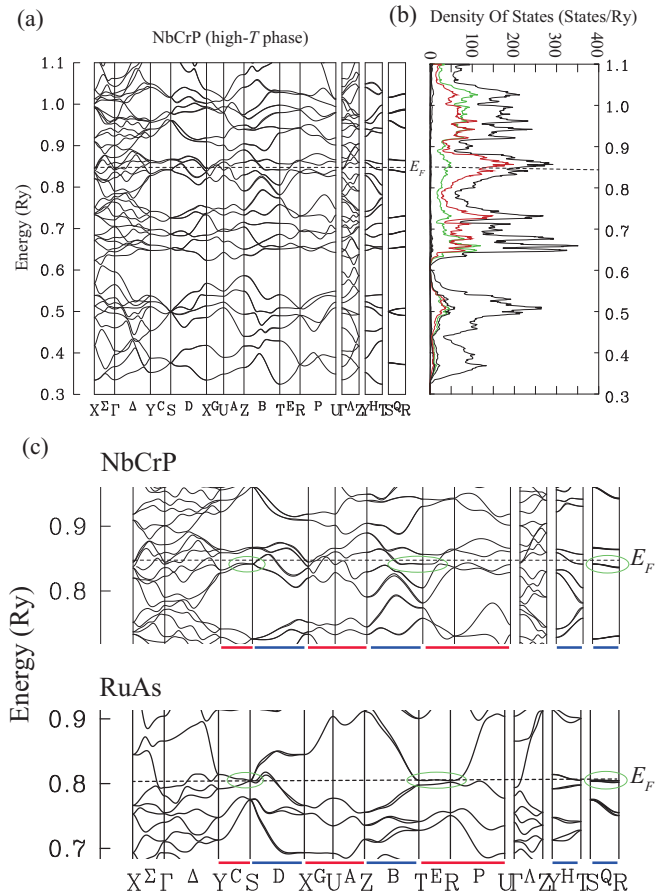}
\caption{\label{fig:band_cal}(a) The energy dispersion and (b) the density of states of NbCrP calculated for orthorhombic TiNiSi-type structure in the $Pnma$ space group in the HT phase. The partial DOS is shown by different colors (red: Cr 3d and green: Nb 4d). 
(c) A comparison of the energy dispersions between NbCrP and RuAs near the Fermi level. 
The four-fold degeneracy is protected along the directions shown by the red lines, and they are slightly lifted along the directions shown by the blue lines due to the spin-orbit coupling. The nearly flat degenerate bands near $E_{\rm F}$ are shown by the green circles. 
}
\end{figure}

Our experimental results show that the first-order phase transition to the nonmagnetic ground state in NbCrP is characterized by the decrease in the density of states at the Fermi level, a symmetry lowering of the crystal, and the loss of the magnetic anisotropy. 
The band structure calculation was performed to check what induces the electronic instability in NbCrP.
Figures \ref{fig:band_cal}(a) and \ref{fig:band_cal}(b) show the energy dispersion and the density of states of NbCrP calculated for an orthorhombic TiNiSi-type structure in the $Pnma$ space group in the HT phase, respectively.
The dispersion and the density of states near the Fermi level ($E_{\rm F}$) originate mainly from Cr-$3d$ orbitals with the secondary contribution from Nb-$4d$ orbitals. 
The $D(E_{\rm F})$ is estimated to be $\sim193$ state/Ry for the primitive cell in the HT phase.
Just below $E_{\rm F}$, the nearly flat bands are partially seen along the Y-S, Z-T, T-R, and S-R axes 
(for the notation for the high symmetry points, see Fig. 9). 
The bands along the Y-S and T-R axes are four-fold degenerate by the nonsymmorphic symmetry of the $Pnma$ space group.
The bands along the Z-T and S-R axes are slightly lifted by the spin-orbit coupling, they are nearly degenerate however.
The band structure of NbCrP shows some similarities to those of  RuAs and RuP, which undergo the nonmagnetic metal-insulator transition, although the overall feature does not resemble them.
Figure \ref{fig:band_cal}(c) shows the band structures for NbCrP and RuAs near $E_{\rm F}$. 
For both systems, the four-fold degeneracy can be seen along the same direction shown by the red lines, and the slight splitting of the bands due to the spin-orbit coupling is also seen along the same direction shown by the blue lines.
The green circles show the nearly flat degenerated bands near $E_{\rm F}$.
The flat bands for RuAs are just on $E_{\rm F}$ \cite{Goto, Kotegawa} (on the other hand, although not shown here, the flat bands for RuP are known to be located above $E_{\rm F}$ by $\sim 0.05$ eV \cite{Goto}). 
In the case of NbCrP, the nearly flat bands are located below $E_{\rm F}$ by $0.006$ Ry $\simeq 0.08$ eV, which is comparable to the case of RuP.
It is also interesting to point out that the similar bands are located below $E_{\rm F}$ by 0.2 eV and the flatness is weakened for RuSb \cite{Goto} in which no phase transition is observed.

\begin{figure}[htb]
\includegraphics[width=0.8\linewidth]{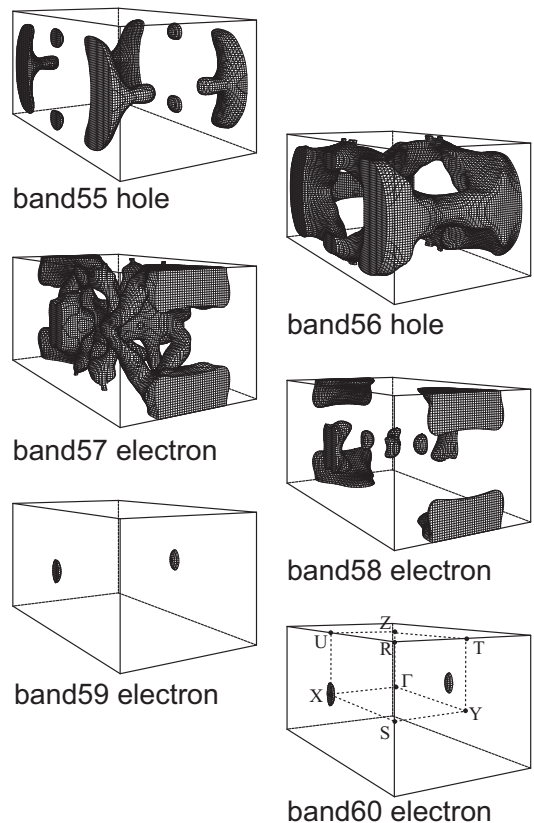}
\caption{\label{fig:band_cal2} Fermi surface of NbCrP calculated for the orthorhombic TiNiSi-type structure in the $Pnma$ space group in the HT phase. They consist several three-dimensional sheets. 
}
\end{figure}

Compared with the case for RuAs, the band dispersion crosses the Fermi level at many points in NbCrP.
This is because it includes contributions from both Cr 3$d$ and Nb 4$d$ orbitals, resulting in a more complicated Fermi surface as shown in Fig. \ref{fig:band_cal2}.
Almost degenerate parts in three pairs of Fermi surfaces ($55-56$ holes, $57-58$ electrons, and $59-60$ electrons) appear on the Brillouin Zone boundary.
This is similar to the case for RuAs \cite{Kotegawa} , however, the Fermi surfaces for NbCrP are more complicated inside the Brillouin Zone. This complexity may cause the remaining of the Fermi surface in LT phase, whereas RuAs is fully gapped in the low temperature insulating phase. 
It is noted that the nesting property can hardly be found in NbCrP.
Therefore, it is unlikely that the phase transition in NbCrP is related to the nesting of the Fermi surface, like the CDW. 

Another important aspect is the density of states at $E_{\rm F}$.
In RuAs and RuP, the Fermi level is located near the peak of the density of states \cite{Goto,Kotegawa}, indicating that the metallic phase is fairly unstable. this is most likely the origin of the disappearance of the Fermi surface with lifting the degeneracy by forming the superlattice structure. 
In NbCrP, however, the Fermi level is located at the local minimum of the density of states, as shown in the Fig. \ref{fig:band_cal}(b), although the sharp peaks are present in the vicinity of $E_{\rm F}$. 
Therefore, it is still puzzling whether or not the phase transition in NbCrP is indeed triggered by the same origin as in RuAs and RuP.
A determination of the crystal structure in the LT phase of NbCrP is essential to elucidate the mechanism of the phase transition.

\section{summary}

We have carried out the x-ray diffraction, electrical resistivity, magnetic susceptibility measurements on the newly synthesized NbCrP crystals to investigate the physical properties.
We also perform $^{31}$P and $^{93}$Nb NMR studies to gain insights into the electronic and magnetic properties of NbCrP from a microscopic point of view.
From the measurements, the nonsymmorphic NbCrP is found for the first time to show the first-order phase transition at $100-150$ K. 
The absence of any trace of magnetic ordering in NMR data indicate that the LT phase is nonmagnetic, while the change in the NMR spectra suggest the lowering of crystal symmetry through the phase transition.
The magnetic susceptibility and NMR data suggest that the spin susceptibility in the LT phase is reduced by $\sim$ 30 \% from that in the HT phase, suggesting that the phase transition is triggered by the electronic instability of the HT phase.
We found that a part of the HT phase can be stabilized down to very low temperatures in the oriented sample.
This indicates that the phase transition is very sensitive to pressure and requires a significant expansion of either the lattice constants or the volume.
The measurement of the oriented sample also revealed the characteristic broad maxima in the temperature dependences of the Knight shift and $1/T_1T$.
This behavior can be interpreted by either the coexistence of FM and AFM interactions or a peak structure of $D(E)$ near the Fermi level.

We also carried out the electronic structure calculations for the HT phase of NbCrP. 
The calculated band structure shows some similarities to those for the nonsymmorphic compound RuAs (with the same space group of $Pnma$) which show the nonmagnetic metal-insulator transition. 
Especially, nearly flat degenerate bands in NbCrP were found to be located just below the Fermi level. 
Although the bands may be related to the phase transition, the Fermi surface instability is not obvious due to the lack of the sharp peak in the density of states at $E_{\rm F}$ in NbCrP. 
Further investigations including a determination of the crystal structure of the LT phase, are desired.
It would also be interesting to investigate the pressure effects on NbCrP to reveal important information about the competition between LT and HT phases at low temperatures.
Understanding the more detailed physical properties of the LT and HT phases is crucial to uncover the origin of the phase transition to the nonmagnetic ground state in NbCrP and the role of the nonsymmorphicity in it.

\begin{acknowledgments}
We thank Khusboo Rana and Yoshinori Haga for experimental supports and helpful discussions.
This work was supported by JSPS KAKENHI Grant Number JP15H05882, JP15H05885, JP15K21732, JP18H04320, and 18H04321 (J-Physics), 15H03689.and 15H05745.
Part of the research was supported by the U.S. Department of Energy, Office of Basic Energy Sciences, Division of Materials Sciences and Engineering. Ames Laboratory is operated for the U.S. Department of Energy by Iowa State University under Contract No. DE-AC02-07CH11358. 
K.Y. also thanks the KAKENHI: J-Physics for financial support that provided an opportunity to visit Ames Laboratory.
\end{acknowledgments}


\begin{references}

\bibitem{burns}
Gerald Burns, Introduction to Group Theory with Application, Allen M. Alper and A. S. Nowick, (ACADEMIC PRESS, New York, 1977).


\bibitem{Hirai}
D. Hirai, T. Takayama, D. Hashizume, and H. Takagi, Phys. Rev. B {\bf 85}, 140509(R) (2012).



\bibitem{Goto}
H. Goto, T. Toriyama, T. Konishi, and Y. Ohta, Physics Procedia {\bf 75}, 91 (2015).


\bibitem{Kotegawa}
H. Kotegawa, K. Takeda, Y. Kuwata, J. Hayashi, H. Tou, H. Sugawara, T. Sakurai, H. Ohta, and H. Harima, Phys. Rev. Mater. {\bf 2}, 055001 (2018).




\bibitem{Sato}
K. Sato, D. Ootsuki, Y. Wakisaka, N. L. Saini, T. Mizokawa, M. Arita, H. Anzai, H. Namatame, M. Taniguchi, D. Hirai, and H. Takagi, arXiv:1205.2669 (2012).



\bibitem{Chen}
R. Y. Chen, Y. G. Shi, P. Zheng, L. Wang, T. Dong, and N. L. Wang, Phys. Rev. B {\bf 91}, 125101 (2015).


\bibitem{Li}
S. Li, Y. Kobayashi, M. Itoh, D. Hirai, and H. Takagi, Phys. Rev. B {\bf 95}, 155137 (2017).


\bibitem{Nakajima}
Y. Nakajima, Z. Mita, H. Watanabe, Y. Ohtsubo, T. Ito, H. Kotegawa, H. Sugawara, H. Tou, and S. Kimura, Phys. Rev. B {\bf 100}, 125151 (2019).

\bibitem{Ootsuki}
D. Ootsuki, K. Sawada, H. Goto, D. Hirai, D. Shibata, M. Kawamoto, A. Yasui, E. Ikenaga, M. Arita, H. Namatame, M. Taniguchi, T. Toriyama, T. Konishi, Y. Ohta, N. L. Saini, T. Yoshida, T. Mizokawa, and H. Takagi, Phys. Rev. B {\bf 101}, 165113 (2019).

\bibitem{Lomnytska}
Y. F. Lomnytska and R. I. Kondratev, Inorganic Materials {\bf 47}, 1072 (2011).

\bibitem{Sheldrick}
G.M. Sheldrick, Acta Crystallogr., Sect. A 64 (2008) 112.

\bibitem{Spek}
A.L. Spek, J. Appl. Crystallogr. 36 (2003) 7.

\bibitem{Kotegawa_CrAs}
H. Kotegawa, S. Nakahara, H. Tou and H. Sugawara, J. Phys. Soc. Jpn. {\bf 83}, 093702 (2014).




\bibitem{Pandey2013} A. Pandey, D. G. Quirinale, W. Jayasekara, A. Sapkota, M. G. Kim, R. S. Dhaka, Y. Lee, T. W. Heitmann, P. W. Stephens, V. Ogloblichev, A. Kreyssig, R. J. McQueeney, A. I. Goldman, A. Kaminski, B. N. Harmon, Y. Furukawa, and D. C. Johnston, Phys. Rev. B {\bf 88}, 014526 (2013).
\bibitem{Wiecki2015} P. Wiecki, V. Ogloblichev, Abhishek Pandey, D. C. Johnston, and Y. Furukawa, Phys. Rev. B {\bf 91}, 220406(R) (2015).
\bibitem{Li2019} Y. Li, Z. Yin, Z. Liu, W. Wang, Z. Xu, Y. Song, L. Tian, Y. Huang, D. Shen, D. L. Abernathy $et~al$., Phys. Rev. Lett. {\bf 122}, 117204 (2019).
\bibitem{Toda2008} M. Toda, H. Sugawara, K. Magishi, T. Saito, K. Koyama, Y. Aoki, and H. Sato, J. Phys. Soc. Jpn. {\bf 77}, 124702 (2008).
\bibitem{Ding2018} Q.-P. Ding, K. Rana, K. Nishine, Y. Kawamura, J. Hayashi, C. Sekine, and Y. Furukawa, Phys. Rev. B {\bf 98}, 155149 (2018).
\bibitem{Moriya1963} T. Moriya, J. Phys. Soc. Jpn. {\bf 18}, 516 (1963).
\bibitem{Narath1968} A. Narath and H. T. Weaver, Phys. Rev. {\bf 175}, 373 (1968)
\bibitem{Slichterbook} C. P. Slichter, Principles~of~Magnetic~Resonance, 3rd ed. (Springer, New York, 1990).



\end{references}
\end{document}